\documentclass[superscriptaddress,twocolumn,showpacs,preprintnumbers,amsmath,amssymb,aps]{revtex4-1}
\usepackage{epsfig}
\pdfoutput=1

\usepackage{subfig}
\usepackage[T1]{fontenc}
\usepackage[latin9]{inputenc}
\usepackage{amsmath,amssymb,amsthm,amsfonts}
\usepackage{ mathrsfs }
\usepackage{slashed}
\usepackage{graphicx}
\usepackage{color,rotating}
\usepackage{dsfont}
\usepackage{setspace}
\usepackage{verbatim}
\usepackage{fancyhdr}
\usepackage{hyperref}

\def\Fig#1{Fig.~\ref{#1}}

\def\s0#1#2{\mbox{\small{$ \frac{#1}{#2} $}}}
\def\0#1#2{\frac{#1}{#2}}

\usepackage{caption}
\newcommand{\beq}{\begin{equation}}
\newcommand{\eeq}{\end{equation}}
\newcommand{\beqa}{\begin{eqnarray}}
\newcommand{\eeqa}{\end{eqnarray}}

\graphicspath{{./figures/}}

\newcommand{\bea}{\begin{eqnarray}}
\newcommand{\eea}{\end{eqnarray}}
\newcommand{\Eq}[1]{Eq.~(\ref{#1})}

\definecolor{darkgreen}{rgb}{0,0.6,0}
\definecolor{gray}{rgb}{.7,.7,.7}

\def\eq#1{(\ref{#1})}
\def\Eq#1{Eq.~(\ref{#1})}
\newcommand {\apgt} {\ {\raise-.5ex\hbox{$\buildrel>\over\sim$}}\ }
\newcommand {\aplt} {\ {\raise-.5ex\hbox{$\buildrel<\over\sim$}}\ }

\def\s0#1#2{\mbox{\small{$ \frac{#1}{#2} $}}}
\def\0#1#2{\frac{#1}{#2}}

\newcommand{\be}{\begin{eqnarray}}
\newcommand{\ee}{\end{eqnarray}}

\begin{document}
\title{Gluon spectral functions and transport coefficients in
  Yang--Mills theory} \vspace{1.5 true cm} 

\author{Michael Haas} \affiliation{Institut f\"ur Theoretische
  Physik, Universit\"at Heidelberg, Philosophenweg 16, 69120
  Heidelberg, Germany} \affiliation{ExtreMe Matter Institute EMMI, GSI
  Helmholtzzentrum f\"ur Schwerionenforschung mbH, Planckstr. 1,
  D-64291 Darmstadt, Germany}

\author{Leonard Fister}
\affiliation{Department for Mathematical Physics, National University
  of Ireland Maynooth, Maynooth, Ireland}

\author{Jan M. Pawlowski} \affiliation{Institut f\"ur Theoretische
  Physik, Universit\"at Heidelberg, Philosophenweg 16, 69120
  Heidelberg, Germany} \affiliation{ExtreMe Matter Institute EMMI, GSI
  Helmholtzzentrum f\"ur Schwerionenforschung mbH, Planckstr. 1,
  D-64291 Darmstadt, Germany}

\begin{abstract}
  We compute non-perturbative gluon spectral functions at finite
  temperature in quenched QCD with the maximum entropy method.  We
  also provide a closed loop equation for the spectral function of the
  energy-momentum tensor in terms of the gluon spectral function. 

  This setup is then used for computing the shear viscosity over
  entropy ratio $\eta/s$ in a temperature range from about $0.4\, T_c$ to
  $4.5\, T_c$. The ratio $\eta /s$ has a minimum at about $1.25\, T_c$
  with the value of about $0.115$. We also discuss extensions of the
  present results to QCD.
\end{abstract}



\maketitle

\noindent {\bf Introduction}  

Heavy-ion collisions at RHIC (Brookhaven) revealed about a decade ago,
that the quark-gluon-plasma (QGP) is well-described by hydrodynamics
\cite{Bellwied:2001cj}.  It was also suggested that the QGP might be
close to exhibit perfect fluidity signaled by a (almost) vanishing
viscosity over entropy ratio $\eta/s$. Since then, many efforts have
been made to increase the insight into the dynamics of the hot plasma,
for a review see \cite{Schafer:2009dj}.

The ratio $\eta/s$ has been conjectured to satisfy a universal lower
bound (KSS-bound) of $1/4\pi$ derived within the AdS-CFT
correspondence \cite{Kovtun:2004de}. Such a minimum can already be
motivated within a quasi-particle picture: there, shear viscosity
relates to a cross section, while the entropy density encodes the
phase space volume of the quasi-particle. In the quasi-particle picture
both quantities are related and their ratio is bounded from below.

Measurements of the elliptic flow variable $v_2$ at RHIC and CERN
indeed indicate a shear viscosity to entropy ratio for the QGP which
is of the order of the AdS-CFT bound
\cite{Shen:2011eg,Agakishiev:2011fs}. In turn, theoretical approaches
to this quantity have to face the problem that perturbation theory is not
applicable in the vicinity of the confinement-deconfinement transition
temperature, and for a strongly correlated plasma.

Transport coefficients can be obtained from the spectral function of
the energy-momentum tensor via the Kubo relations
\cite{Kubo:1957}. However, most non-perturbative methods such as
lattice QCD and functional continuum methods are so far limited to the
computation of Euclidean correlation functions of the energy-momentum
tensor, see e.g.\ \cite{Karsch:1986cq, Meyer:2007ic,
  Meyer:2007dy,Suganuma:2009zs} for lattice results. The related
spectral function is then obtained via an integral equation. The
latter has to be inverted from a discrete set of points, or more
generally from numerical data, for example with the maximum entropy
method (MEM), e.g.\ \cite{Asakawa:2000tr} or the Tikhonov regularization, 
e.g.\ \cite{Dudal:2013vf}. So far, the resulting spectral functions 
$\rho(\omega,\vec{p}) $ are subject to large statistical as well as 
systematical errors.

In principle, MEM and similar inversion methods are powerful tools for
providing reliable spectral functions, but this requires accurate
initial Euclidean correlation functions and some knowledge about their
real-time asymptotics and complex structure. Whether such a situation
applies directly to the correlation function of the energy-momentum
tensor is difficult to answer and is part of the systematic
error. 

In the present work we apply MEM for the computation of the gluon
spectral functions in Landau gauge Yang--Mills theory at finite
temperature from Euclidean propagators obtained from FRG calculations
\cite{Fister:2011uw}. It is well known that the gluon spectral
function exhibits positivity violation, e.g.\ \cite{Maas:2005ym}, and
we implement an adjustment of MEM for non-positive
functions. We provide gluonic spectral functions
for $0.4\, T_c \lesssim T\lesssim4.5\, T_c$. Moreover, the zero
temperature extrapolation of our results agrees well with the direct $T=0$ 
computation with Dyson--Schwinger equations in \cite{Strauss:2012dg}.

The gluon spectral functions are then used to compute the viscosity
over entropy ratio in this temperature range from a compact closed
expression of the spectral function of the energy-momentum tensor in
terms of gluon propagators and classical and full vertices. \\[-2ex]

\noindent {\bf Maximum Entropy Method}

The spectral function $\rho(\omega,\vec{p}\,)$ is related to the
Euclidean propagator $G(i\omega_n,\vec{p}\,)$ via the integral equation 
\begin{equation}
\label{eq:MEM_main}
 G(\tau,\vec{p}\,)=\int\limits_0^\infty\frac{d\omega}{2\pi} 
K_T(\tau,\omega)\rho(\omega,\vec{p}\,)\,,
\end{equation}
with 
\begin{equation} 
\label{eq:kernel} 
K_T(\tau,\omega)=(1+n(\omega))e^{-\omega\tau}+
n(\omega)e^{\omega\tau}\,, 
\end{equation} 
with thermal distribution $n(\omega)=1/(e^{\omega/T}-1)$. In
\eq{eq:MEM_main}, $G(\tau,\vec{p}\,)$ denotes the Fourier transform of
the Euclidean propagator $G(i\omega_n,\vec{p}\,)$ in a slight abuse of
notation. 

The inversion of \eq{eq:MEM_main} is not unique, and it is necessary
to include information on the general shape of the spectral
function. That is achieved by introducing a positive model function
$m(\omega,\vec{p}\,)$, containing all available information on the
asymptotic behavior (shape) of $\rho(\omega,\vec{p}\,)$.  In practice
$m(\omega,\vec{p}\,)$ encodes the correct UV behavior known from
perturbation theory. MEM minimizes the quantity
$Q(\vec{p}\,)=L(\vec{p}\,)-\alpha S(\vec{p}\,)$ with
 \begin{equation}
\label{eq:likelihood}
L(\vec{p}\,)=\frac12\int\limits_0^{\frac\beta2}\frac{d\tau}{\sigma^2
(\tau,\vec{p}\,)}(G(\tau,\vec{p}\,)-G_{\rho}(\tau,\vec{p}\,))^2\,,
\end{equation}
and 
\begin{equation}
  \label{eq:entropy}
  S(\vec{p}\,)=\int_0^\infty d\omega [\rho(\omega,\vec{p}\,)-m(\omega,
  \vec{p}\,)-\rho(\omega,\vec{p}\,)\log{\frac{\rho(\omega,\vec{p}\,)}{
m(\omega,\vec{p}\,)}}]\,.
\end{equation}
Here $\sigma(\tau,\vec{p}\,)$ encodes the uncertainties of the input
correlator. In \eq{eq:likelihood} $G_{\rho}(\tau,\vec{p}\,)$ denotes
the propagator calculated from the MEM spectral function via
\eq{eq:MEM_main}.  The likelihood term \eq{eq:likelihood} and the
entropy term \eq{eq:entropy} of Shannon-Jaynes type
\cite{Jaynes:1957zza} are related by the weight parameter $\alpha$.
The weight parameter regulates the relative importance of the model
with respect to the correlator, and can be integrated out, see e.g.\
\cite{Asakawa:2000tr}.

For positive model functions, the MEM ansatz for the spectral function
is intrinsically positive. However, the gluon spectral functions
$\rho(\omega,\vec p\,)$ show positivity violation for large
frequencies and sufficiently low momenta. This property relates to the
fact that the gluon is no asymptotic state, and is taken into account
by parameterizing the spectral functions as a difference of two
positive model functions: $\rho(\omega,\vec{p}\,)=
\rho_s(\omega,\vec{p}\,)- s(\omega,\vec{p}\,)$. Such splittings have
been also used for, e.g., quark spectral functions, for recent work see 
\cite{Qin:2013ufa}. The shift function
$s(\omega,\vec{p}\,)$ should allow for a finite violation of
positivity, i.e.\ no poles and essential singularities. The propagator
corresponding to $\rho_s(\omega,\vec{p}\,)$ is 
\begin{equation}
\label{eq:shifted_propagator}
G_s(\tau,\vec{p}\,)=G(\tau,\vec{p}\,)+\Delta G(\tau,\vec{p}\,)
\end{equation}
where
\begin{equation}
\label{eq:Delta_G}
\Delta G(\tau,\vec{p}\,)=\int\limits_0^\infty\frac{d\omega}{2\pi}
K_T(\tau,\omega)s(\omega,\vec{p}\,)\,.
\end{equation}
The finiteness of the integral \eq{eq:Delta_G} is guaranteed by the
known, perturbative, asymptotic behavior of $\rho(\omega,\vec p\,)$.\\[-2ex]

\noindent {\bf Viscosity}

One of the main goals of the present work is the computation of the
viscosity over entropy ratio $\eta/s$ as a function of temperature in
the vicinity of the phase transition temperature. With the Kubo
relation the shear viscosity $\eta$ is computed from the slope of
the spectral function $\rho_{\pi\pi}$ of the spatial, traceless part
of the energy-momentum tensor $\pi_{ij}$ at vanishing frequency,
\begin{equation}
\label{eq:Kubo}
\eta=\lim\limits_{\omega \to 0} \0{1}{20}\frac {\rho_{\pi\pi}(\omega,\vec{0}\,)}{\,\omega}\,,
\end{equation}
with
\begin{equation}
\label{eq:EMT}
\rho_{\pi\pi}(\omega,\vec{p}\,)= \int \frac {d x_0}{2\pi} 
\int \frac {d^3 x} {(2\pi)^3}  
e^{-i\omega x_0+i\vec{p}\vec{x}}\langle[\pi_{ij}(x),\pi_{ij}(0)]\rangle\,.
\end{equation}
For the computation of \eq{eq:EMT} we use that general correlation
functions can be written in terms of propagators
and field derivatives, see e.g.\ \cite{Pawlowski:2005xe},
\begin{equation}\label{eq:looprep}
  \langle \pi_{ij}[A] \pi_{ij}[A]\rangle = \pi_{ij}[ G_{A \phi_i}\cdot 
  \0{\delta}{\delta \phi_i}
  +\bar A]\, \pi_{ij}[ G_{A \phi_i}\cdot \0{\delta}{\delta \phi_i}+\bar A]\,, 
\end{equation}
where $\phi=(\bar A,C,\bar C)$ stands for the expectation values of
the fields, e.g.\ $\bar A=\langle A\rangle$, and $G_{\phi_i \phi_j}=
\langle \phi_i\,\phi_j\rangle - \langle \phi_i\rangle \,\langle \phi_j\rangle$
is the propagator of the respective fields. 

\begin{figure}%
  \subfloat[The three different classes of diagrams: two
  energy-momentum tensors (double lines) connected by 2,3,4 full
  internal gluon propagators.]%
  {%
    \makebox[\columnwidth]{%
                \includegraphics[width=.4\textwidth]{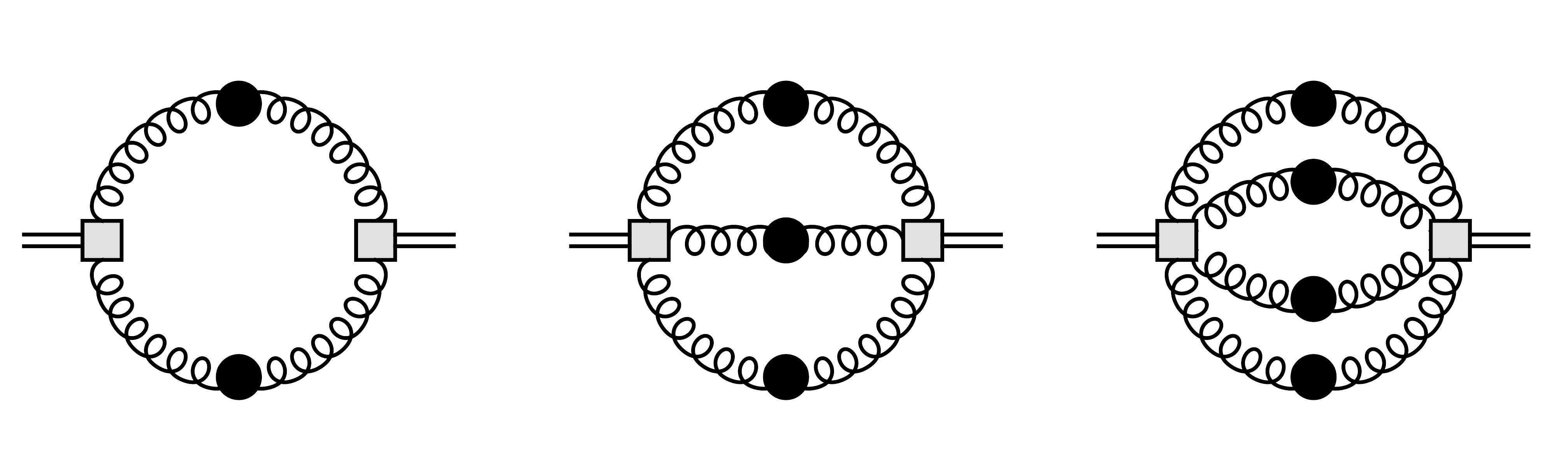}%
                 \label{fig:diagrams}%
               }%
             }\hfill%
             \\%
             \subfloat[Examples for effective energy-momentum tensor vertex corrections for the one-loop diagram in
             \Fig{fig:diagrams}.]%
             {%
               \makebox[\columnwidth]{%

                 \includegraphics[width=.45\textwidth]{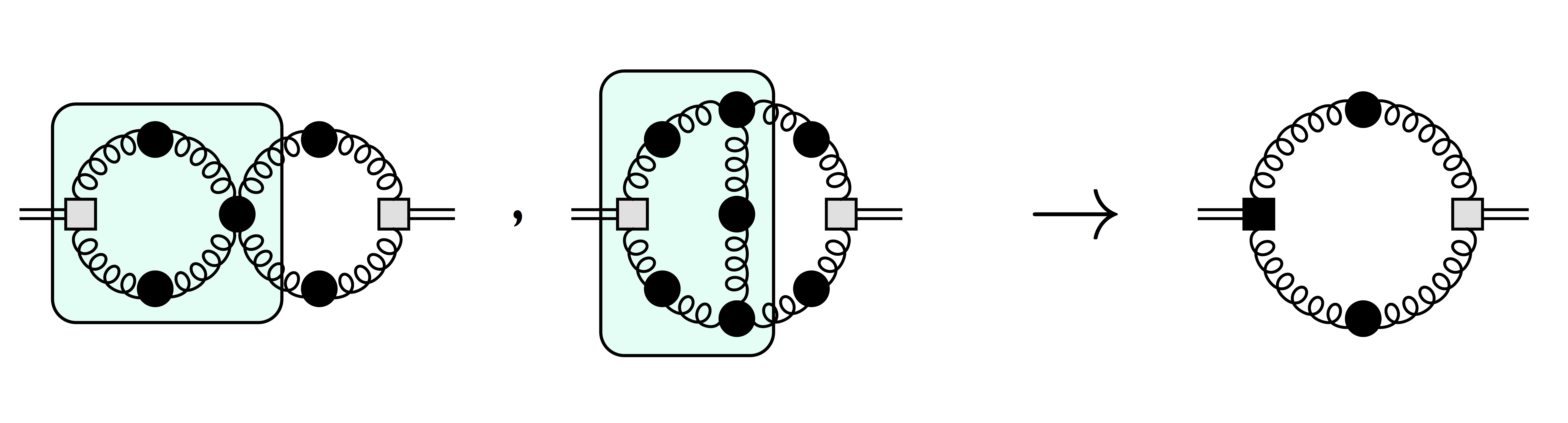}%
      	    \label{fig:vertex_correction}%
        }%
    }\hfill%
    \caption{Diagrams contributing to the energy-momentum tensor
      spectral function.}%
    \label{fig:approx}%
    \end{figure}

\Eq{eq:looprep} consists of a finite number of connected diagrams in
full propagators. The one-particle irreducible diagrams can be divided
into two classes. The first class consists of one- to three-loop
diagrams with gluon propagators that simply connect one $\pi_{ij}$
with the other, see \Fig{fig:diagrams}. The second class consists of
diagrams that can be interpreted as effective vertex corrections of
the first class. A simple example is depicted in
\Fig{fig:vertex_correction}, the full diagrammatics will be discussed
elsewhere \cite{CHPS}.

    In the present work we concentrate on temperatures of the order of
    $T_c$. In \cite{Fister:2013bh} it has been discussed in the
    context of the Polyakov loop potential that higher loop
    corrections in such an expansion in full propagators and full and
    classical vertices can be minimized within an optimized RG--scheme
    for temperatures about $T_c$. Indeed, the explicit computation
    confirms that higher loop orders in \Fig{fig:approx} are
    suppressed at these temperatures \cite{CHPS}.  Accordingly, the
    weighted difference of the full computation of the Polyakov loop potential
    and the one loop computation in full propagators can be used as an
    estimate for the systematic error.

In conclusion, for temperatures about $T_c$ we can restrict ourselves
to the one-loop contribution in \Fig{fig:diagrams}.  Note also, that
connected contributions are of higher order and hence are dropped in
the present computation. In this approximation the spectral function
$\rho_{\pi\pi}$ reads
\begin{eqnarray}\nonumber 
\rho_{\pi\pi}(p)&=&\text{Im}\Bigl[\int\0{d^4 k}{(2 \pi)^4}
\pi^{(2)}(k,p+k)G(p+k)\\
&& \hspace{1.65cm} \times \pi^{(2)}(p+k,k)G(k)\Bigr]\,,
\label{eq:one_loop_EMT}\end{eqnarray}
where $\pi^{(2)}$ denotes the two-gluon vertex of the energy-momentum
tensor $\pi$, and $p=(\omega,\vec{0}\,)$. For the sake of brevity we
omitted the colour and Lorentz indices. The gluon propagators in
Landau gauge at finite temperature have two separate tensor
structures, longitudinal and transverse to the heat bath. For each
tensor structure we have a scalar propagator $G_{L/T}(p_0^2,\vec
  p^{\,2})$. In order to evaluate \eqref{eq:one_loop_EMT} we insert the
tensor expression for $G(p)$ and use the cutting rules within the
real-time formalism for the scalar parts of the propagators.  Finally,
we insert the spectral representation for the propagators.  The
details will be discussed in \cite{CHPS} and we only present the
results,
\begin{eqnarray}\nonumber 
  \rho_{\pi\pi}(\omega)&=&\frac{2d_A}{3}\int
  \frac{d^4k}{(2\pi)^4}[n(k^0)-n(k^0+\omega)]\\\nonumber 
  &&\times\{V_1(k,\omega)\rho_T(k^0,\vec{k}\,)
  \rho_T(k^0+\omega,\vec{k}\,)\\\nonumber 
  &&+V_2(k,\omega)\rho_T(k^0,\vec{k}\,)
  \rho_L(k^0+\omega,\vec{k}\,)\\
  &&+V_3(k,\omega)\rho_L(k^0,\vec{k}\,)
  \rho_L(k^0+\omega,\vec{k}\,)\}\,,
\label{eq:rho}\end{eqnarray}
with $d_A=N_c^2-1$, and 
\begin{eqnarray}\nonumber 
 V_1(k)&=&7(k^2)^2-10k_0^2\vec{k}\,^2+7k_0^4\\\nonumber 
 V_2(k)&=&6k_0^2(k_0^2-\vec{k}\,^2)\\
 V_3(k)&=&2(k_0^2-\vec{k}\,^2)^2\,. 
\label{eq:Vis}\end{eqnarray}
The $V_i$'s are the coefficients in Landau gauge arising from the
vertex contractions. If we apply the present cutting-rule approach to
Coloumb gauge, the coefficients agree with that obtained in
\cite{Aarts:2002cc} from a Matsubara approach. 

Equation \eq{eq:rho} has an important feature. When taking the
derivative with respect to $\omega$ at $\omega=0$, the derivative only
hits the thermal distribution function. Thus, the results for the
viscosity are not sensitive to the slope of the gluon spectral
functions at vanishing frequency. Differentiating \eq{eq:EMT} with
respect to $\omega$ at $\omega=0$ yields
\begin{eqnarray}\nonumber 
  \eta&=-\frac{2d_A}{3}\int \frac{d^4k}{(2\pi)^4}n'(k^0)\times\{V_1(k)
\rho_T^2(k^0,\vec{k}\,)\\[1ex]
  &+V_2(k)\rho_T(k^0,\vec{k}\,)\rho_L(k^0,\vec{k}\,)+V_3(k)
\rho_L^2(k^0,\vec{k}\,)\}\,.
\label{eq:zero_omega}
\end{eqnarray}
For the computation of the viscosity over entropy ratio $\eta/s$ we
take the entropy obtained within lattice calculations
in \cite{Karsch:2001cy,Meyer:2009tq,Borsanyi:2012ve}.\\[-2ex]

\noindent {\bf Results}

The results are computed from the Euclidean Yang--Mills propagators
at vanishing frequency $\omega_n=0$ for the longitudinal and transverse
gluons for different temperatures obtained by FRG techniques,
\cite{Fister:2011uw}. For comparison and error estimates we also
utilize lattice results from \cite{Fischer:2010fx, Maas:2011se,
  Maas:2011ez}, see also \cite{Cucchieri:2007ta, Cucchieri:2011di,
  Aouane:2011fv, Cucchieri:2012nx}. We have done this comparison for
the temperature regime at about $T_c$ where the respective results
agree well.

In our computations we have approximated the higher Matsubara modes in
  the scalar propagators $G_{L/T}$ for $\omega_n\neq 0$ with
  $G_{L/T}(\omega_n^2,\vec{p}^{\,2}\,)=G_{L/T}(0,\omega_n^2+{\vec{p}}^{\,2})$. This
  is a quantitative approximation with a small error margin of $<1\%$,
  see \cite{Fister:2011uw}, well below the systematic errors in the
  present computation to be discussed later.

\begin{figure}[t]%
     \subfloat[$T= 0.79\,T_c$.]{
\includegraphics[width=0.24\textwidth]{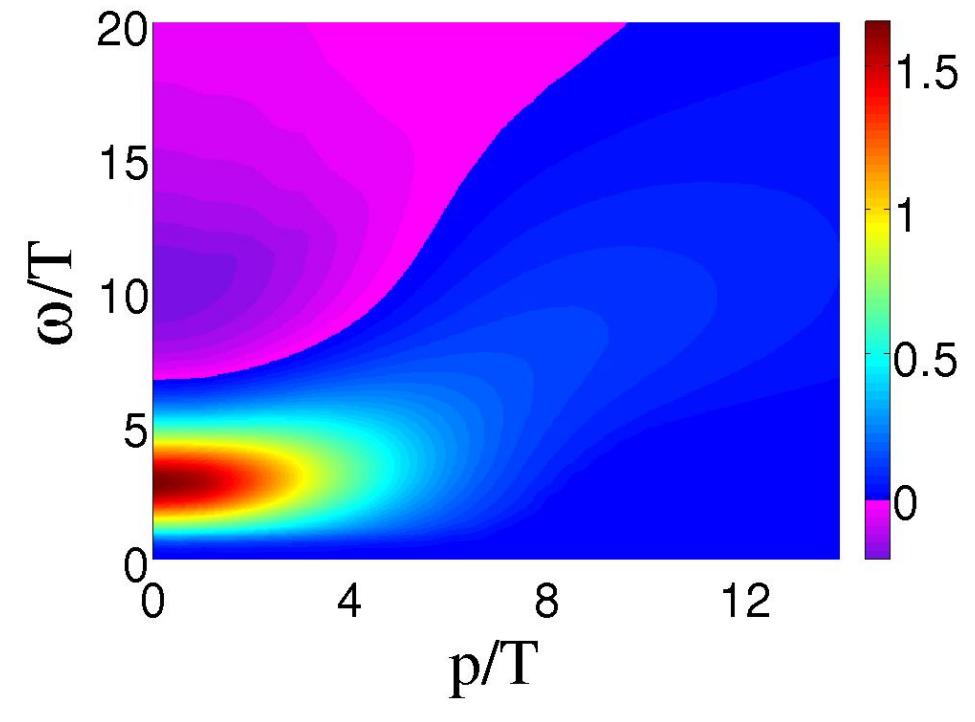}}%
     \subfloat[$T=1.59\,T_c$.]{
\includegraphics[width=0.24\textwidth]{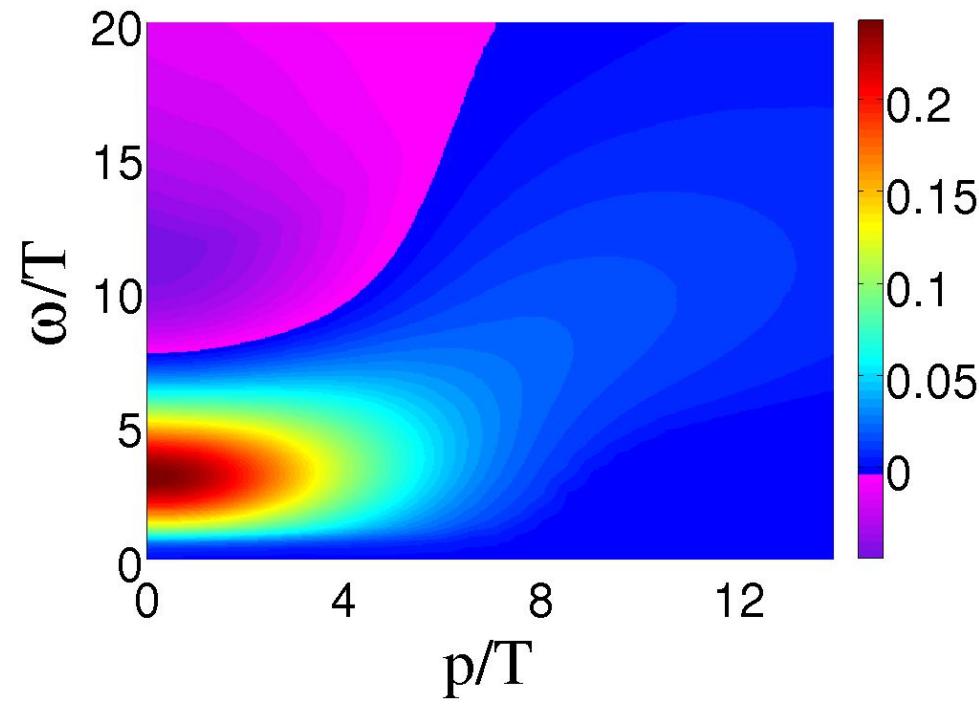}}%
     \\%
     \subfloat[$T=2.77\,T_c$.]{
\includegraphics[width=0.24\textwidth]{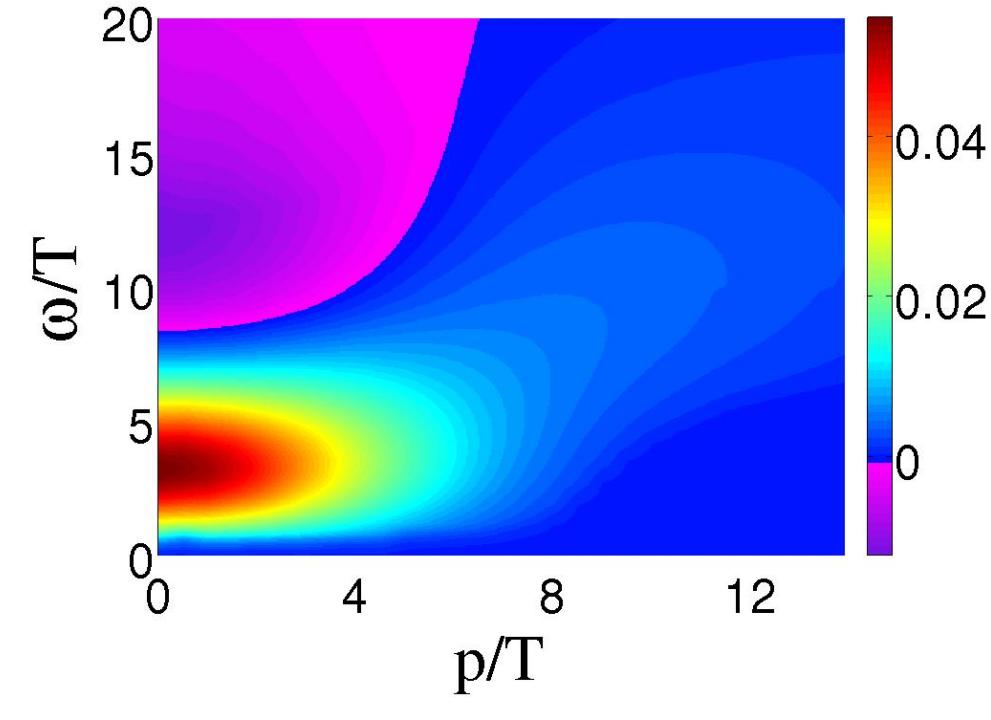}}%
     \subfloat[$T=3.96\,T_c$.]{
\includegraphics[width=0.24\textwidth]{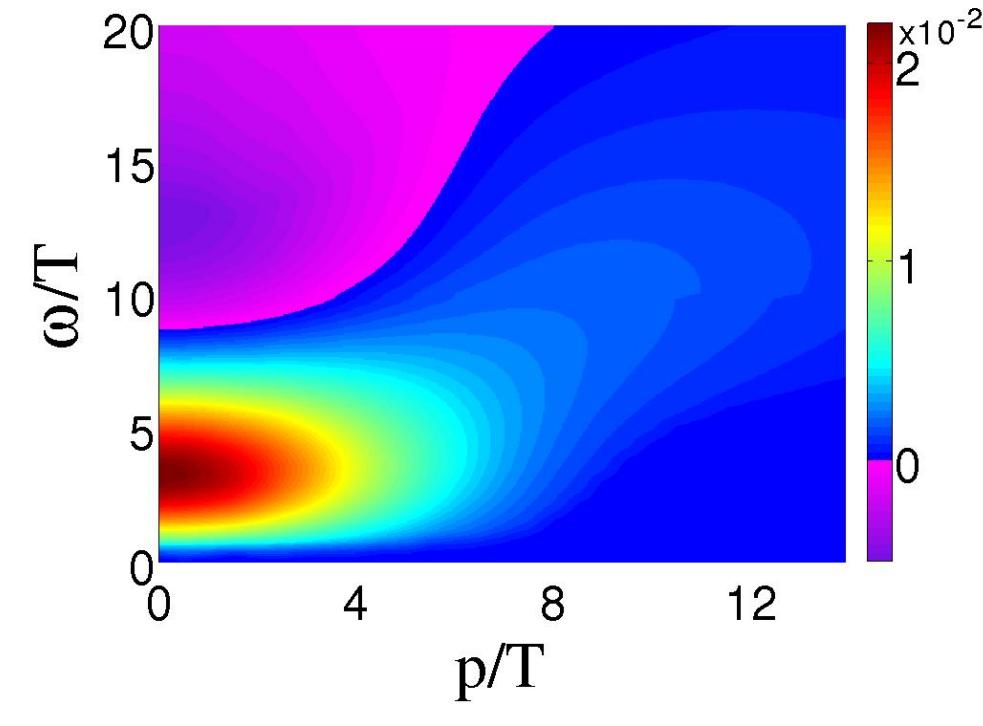}}%
  \caption{Thermal dependence of transverse gluon spectrum.}%
  \label{fig:spectrum_T_gluons}%
\end{figure}

\begin{figure}[t]%
  \includegraphics[width=0.45\textwidth]{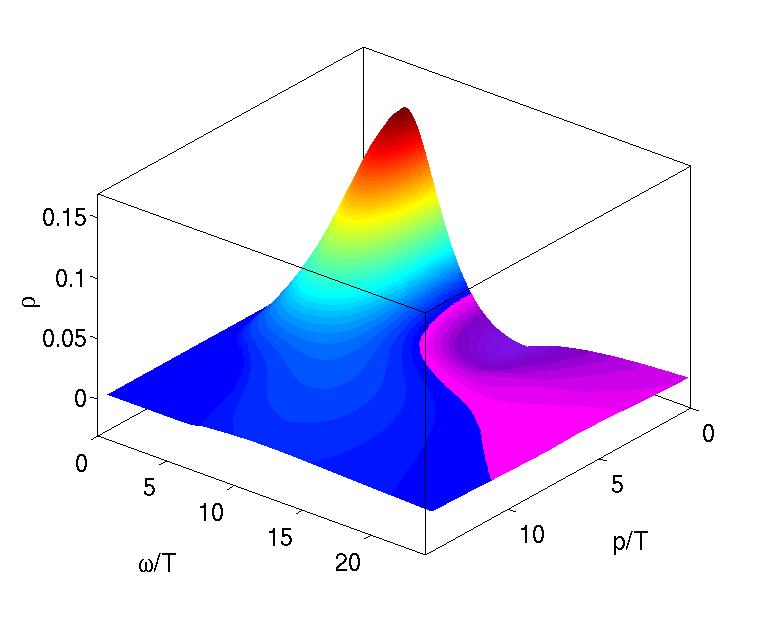}%
  \caption{Transverse gluon spectral function $\rho(\omega,\vec{p}\,)$
    for $T=1.98\,T_c$.}%
      \label{fig:contour_diff_view}%
\end{figure}

\Fig{fig:spectrum_T_gluons} shows the MEM-results for the transverse
gluon spectral functions for a temperature range from $T=0.79\,T_c$ to
$T=3.96\,T_c$. The common features of all calculated gluonic spectral
functions are a broad maximum at $\omega/T\approx2.0-3.0$ and a
violation of positivity at low spatial momenta. At larger momenta, the
peak smears out and approaches the line $\omega=p$, see also
\Fig{fig:contour_diff_view} for the transverse spectral function at
$T=1.98\,T_c$. The fact that the (positive) peak position for fixed
$\omega$ as a function of $p$ is stationary for $p\lesssim$ 6 seems to
be due to the negativity of the spectral function, which inhibits the
bending of peak towards the main diagonal. Hence the gluon spectral
functions do not show the characteristic diagonal structure of
quasi-particle spectral functions, see \Fig{fig:contour_diff_view}.
Such a quasi-particle picture has been used e.g.\ in
\cite{Peshier:2004bv, PC_Frankfurt} where the model gluon spectral
functions have sharper peaks with comparable peak heights.

With increasing temperature the peak broadens slightly more than
linearly in $T$, while the area under the peak remains approximately
constant. In the limit $T\to 0$ this would lead to a delta-like peak,
as seen in \cite{Strauss:2012dg}. The magnitude of the minimum is
about $10\%$ of the maximum of the spectral function.

The dependence of our results on the shift function is surprisingly
small. We have employed different ans\"{a}tze and found that only the
necessary condition, that the shift function must be larger in
magnitude than the negative values of the spectral function at the
respective point, must be fulfilled. All other features could be
chosen freely as long as \eq{eq:Delta_G} is kept finite. The
longitudinal spectral functions show no different behavior and differ
only slightly from the transverse spectral functions.

\begin{figure}[t]%
              \includegraphics[width=0.49\textwidth]{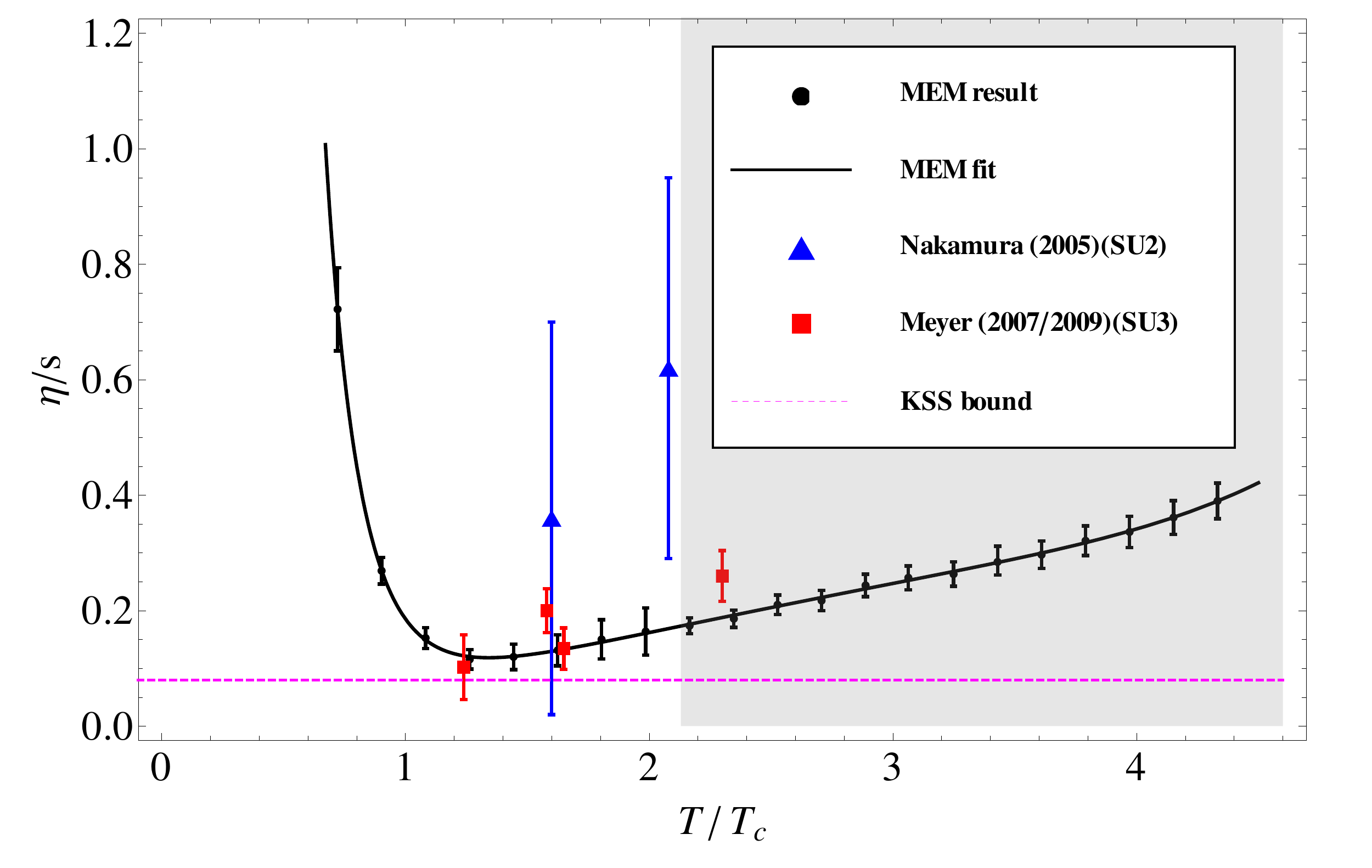}%
              \caption{Viscosity over entropy ratio $\eta/s$ for SU(3)
                gauge theory. The AdS/CFT bound is displayed, as well
                as lattice results from
                \cite{Meyer:2007dy,Meyer:2009jp,Nakamura:2005yf}. The black
                error bars indicate the combined systematic error from
                both MEM computation and one-loop approximation as
                discussed above. In the shaded region only MEM errors are
		displayed.}%
      \label{fig:visc_over_entropy}%
\end{figure}

We have calculated the shear viscosity applying the Kubo relation
\eq{eq:Kubo} for the spectral function of the energy-momentum tensor
at vanishing frequency and divided by the entropy density. In
\Fig{fig:visc_over_entropy} we show the results as a function of
temperature. The black error bars indicate the combined systematic
error from both MEM computation and one-loop approximation as
discussed above. In the shaded region only MEM errors are displayed.
The error analysis exhibits a small systematic and statistical error
for temperatures $T_c \leq T\lesssim 2 T_c$. For larger temperature
the one-loop approximation within the optimized RG-scheme becomes
worse and higher order diagrams have to be included. In turn, for
smaller temperatures $T \leq T_c$ additionally the accuracy of the
spectral functions has to be increased in order to provide reliable
quantitative results. Moreover, the uncertainty in the relative
temperature scales on the lattice and the functional methods gets
important due to the strong temperature dependence in this regime. An
additional systematic error relates to the systematic error of the
input data. We have also computed the ratio $\eta/s$ from the lattice
propagators in \cite{Fischer:2010fx} for temperatures about $T_c$ and
the result varies with about $5\%$. This error is not included in the plot 
 in \Fig{fig:visc_over_entropy}. 

The curve in \Fig{fig:visc_over_entropy} exhibits a clear minimum at
$T= 1.25 T_c$ with a value of $\eta/s=0.115(17)$. This region is well
in the regime with small systematic and statistical errors. Below the
critical temperature we find a steep rise of $\eta/s$ towards lower
temperatures due to the decrease of the entropy density. In view of
the above error analysis this should be seen as a qualitative
result. Within the present accuracy we also cannot resolve potential
signatures of the first order phase transition. Our results agree
qualitatively with model computations of $\eta/s$, see
e.g. \cite{Marty:2013ita}. Note also, that for $T \leq T_c$ glueballs
are expected to be the relevant degrees of freedom. It would be
interesting to see how the present results fit into a corresponding
quasi-particle picture based on glueballs.
\\[-2ex]

\noindent {\bf Conclusions}

We have computed gluon spectral functions from non-perturbative,
Euclidean propagators in Landau gauge finite temperature Yang--Mills
theory. This has been done with a modified version of the maximum
entropy method that allows for negative parts in the spectral
functions.  As expected the spectral functions show a violation of
positivity. Our results cover the temperature regime $0.4\, T_c
\lesssim T\lesssim4.5\, T_c$. We have computed the shear viscosity 
$\eta$ from a closed expression in terms of the gluon spectral 
function. With the lattice entropy taken from 
\cite{Karsch:2001cy,Meyer:2009tq,Borsanyi:2012ve} this leads us
to the viscosity over entropy ratio $\eta/s$ in the above temperature
range.  We find a minimum value of $\eta/s=0.115(17)$ at $T= 1.25
T_c$ which is close to, but above the KSS bound of
$\eta/s=1/(4\pi)$. Interestingly, the results agree within the errors
with previous lattice computations,
\cite{Meyer:2007dy,Meyer:2009jp}. Given the very different
computational methods, this provides non-trivial support for the
respective results. In \cite{Meyer:2009jp} a mapping of Yang--Mills
$\eta/s$ to QCD is proposed for $T=2.3T_c$. Adapting the procedure we
propose a minimal $\eta/s$ for QCD of 0.18. The present framework is
readily extended to full QCD with dynamical fermions, which is
currently
under investigation. \\[-2ex]

\noindent {\bf Acknowledgements} \\[0ex]
We thank N.~Christiansen and N.~Strodthoff for discussions and work on
related subjects.  This work is supported by the Helmholtz Alliance
HA216/EMMI and by ERC-AdG-290623. LF is supported by the Science
Foundation Ireland in respect of the Research Project
11-RFP.1-PHY3193, MH acknowledges support by the
Landesgraduiertenf\"orderung Baden-W\"urttemberg via the Research
Training Group ``Quantum Many-body Dynamics and Nonequilibrium
Physics''.

\bibliographystyle{bibstyle}
\bibliography{viscosity}

\begin{thebibliography}{10}

\bibitem{Bellwied:2001cj}
STAR collaboration, R.~Bellwied,
\newblock p. 132 (2001), hep-ph/0112250.

\bibitem{Schafer:2009dj}
T.~Schafer and D.~Teaney,
\newblock Rept.Prog.Phys. {\bf 72}, 126001 (2009), 0904.3107.

\bibitem{Kovtun:2004de}
P.~Kovtun, D.~Son, and A.~Starinets,
\newblock Phys.Rev.Lett. {\bf 94}, 111601 (2005), hep-th/0405231.

\bibitem{Shen:2011eg}
C.~Shen, U.~Heinz, P.~Huovinen, and H.~Song,
\newblock Phys.Rev. {\bf C84}, 044903 (2011), 1105.3226.

\bibitem{Agakishiev:2011fs}
STAR Collaboration, H.~Agakishiev {\em et~al.},
\newblock Phys.Lett. {\bf B704}, 467 (2011), 1106.4334.

\bibitem{Kubo:1957}
R.~Kubo,
\newblock Journal of the Physical Society of Japan {\bf 12}, 570 (1957).

\bibitem{Karsch:1986cq}
F.~Karsch and H.~Wyld,
\newblock Phys.Rev. {\bf D35}, 2518 (1987).

\bibitem{Meyer:2007ic}
H.~B. Meyer,
\newblock Phys.Rev. {\bf D76}, 101701 (2007), 0704.1801.

\bibitem{Meyer:2007dy}
H.~B. Meyer,
\newblock Phys.Rev.Lett. {\bf 100}, 162001 (2008), 0710.3717.

\bibitem{Suganuma:2009zs}
H.~Suganuma, T.~Iritani, A.~Yamamoto, and H.~Iida,
\newblock PoS {\bf QCD-TNT09}, 044 (2009), 0912.0437.

\bibitem{Asakawa:2000tr}
M.~Asakawa, T.~Hatsuda, and Y.~Nakahara,
\newblock Prog.Part.Nucl.Phys. {\bf 46}, 459 (2001), hep-lat/0011040.

\bibitem{Dudal:2013vf}
D.~Dudal, P.~J. Silva, and O.~Oliveira,
\newblock PoS {\bf ConfinementX}, 033 (2012), 1301.2971.

\bibitem{Fister:2011uw}
L.~Fister and J.~M. Pawlowski,
\newblock (2011), 1112.5440.

\bibitem{Maas:2005ym}
A.~Maas,
\newblock Mod. Phys. Lett. {\bf A20}, 1797 (2005), hep-ph/0506066.

\bibitem{Strauss:2012dg}
S.~Strauss, C.~S. Fischer, and C.~Kellermann,
\newblock Phys.Rev.Lett. {\bf 109}, 252001 (2012), 1208.6239.

\bibitem{Jaynes:1957zza}
E.~Jaynes,
\newblock Phys.Rev. {\bf 106}, 620 (1957).

\bibitem{Qin:2013ufa}
S.-x. Qin and D.~H. Rischke,
\newblock (2013), 1304.6547.

\bibitem{Pawlowski:2005xe}
J.~M. Pawlowski,
\newblock Annals Phys. {\bf 322}, 2831 (2007), hep-th/0512261.

\bibitem{CHPS}
N.~Christiansen, M.~Haas, J.~M. Pawlowski, and N.~Strodthoff,
\newblock in preparation  (2013).

\bibitem{Fister:2013bh}
L.~Fister and J.~M. Pawlowski,
\newblock (2013), 1301.4163.

\bibitem{Aarts:2002cc}
G.~Aarts and J.~M. {Martinez Resco},
\newblock JHEP {\bf 0204}, 053 (2002), hep-ph/0203177.

\bibitem{Karsch:2001cy}
F.~Karsch,
\newblock Lect. Notes Phys. {\bf 583}, 209 (2002), hep-lat/0106019.

\bibitem{Meyer:2009tq}
H.~B. Meyer,
\newblock Phys.Rev. {\bf D80}, 051502 (2009), 0905.4229.

\bibitem{Borsanyi:2012ve}
S.~Borsanyi, G.~Endrodi, Z.~Fodor, S.~Katz, and K.~Szabo,
\newblock JHEP {\bf 1207}, 056 (2012), 1204.6184.

\bibitem{Fischer:2010fx}
C.~S. Fischer, A.~Maas, and J.~A. M{\"u}ller,
\newblock Eur. Phys. J. {\bf C68}, 165 (2010), 1003.1960.

\bibitem{Maas:2011se}
A.~Maas,
\newblock Phys.Rept. {\bf 524}, 203 (2013), 1106.3942.

\bibitem{Maas:2011ez}
A.~Maas, J.~M. Pawlowski, L.~von Smekal, and D.~Spielmann,
\newblock Phys.Rev. {\bf D85}, 034037 (2012), 1110.6340.

\bibitem{Cucchieri:2007ta}
A.~Cucchieri, A.~Maas, and T.~Mendes,
\newblock Phys. Rev. {\bf D75}, 076003 (2007), hep-lat/0702022.

\bibitem{Cucchieri:2011di}
A.~Cucchieri and T.~Mendes,
\newblock (2011), 1105.0176.

\bibitem{Aouane:2011fv}
R.~Aouane {\em et~al.},
\newblock (2011), 1108.1735.

\bibitem{Cucchieri:2012nx}
A.~Cucchieri and T.~Mendes,
\newblock PoS {\bf LATTICE2011}, 206 (2011), 1201.6086.

\bibitem{Peshier:2004bv}
A.~Peshier,
\newblock Phys.Rev. {\bf D70}, 034016 (2004), hep-ph/0403225.

\bibitem{PC_Frankfurt}
H.~Berrehrah,
\newblock for the PHSD group, private communication\!\! .

\bibitem{Meyer:2009jp}
H.~B. Meyer,
\newblock Nucl.Phys. {\bf A830}, 641C (2009), 0907.4095.

\bibitem{Nakamura:2005yf}
A.~Nakamura and S.~Sakai,
\newblock Nucl.Phys. {\bf A774}, 775 (2006), hep-lat/0510039.

\bibitem{Marty:2013ita}
R.~Marty, E.~Bratkovskaya, W.~Cassing, J.~Aichelin, and H.~Berrehrah,
\newblock (2013), 1305.7180.

\end{thebibliography}

\end{document}